\documentstyle{mn}
\input{epsf}
\input{rotate}

\title[Properties of galaxies and clusters]{Properties of
spherical galaxies and clusters with an NFW density profile}

\author[Ewa L. {\L}okas and Gary A. Mamon]{Ewa L. {\L}okas$^1$ and Gary A.
    Mamon$^{2,3}$\\ $^1$Copernicus Astronomical Center, Bartycka 18,
    00--716 Warsaw, Poland\\ $^2$Institut d'Astrophysique de Paris
   (CNRS UPR 341), 98 bis Bd Arago, F--75014 Paris, France \\ $^3$DAEC
    (CNRS UMR 8631), Observatoire de Paris,
    Place Jules Janssen, F--92195 Meudon, France}

\begin{document}

\maketitle

\begin{abstract}
Using the standard dynamical theory of spherical systems, we
calculate the properties of spherical galaxies and clusters whose density
profiles obey the universal form first obtained in high resolution
cosmological $N$-body simulations by Navarro, Frenk \& White.
We adopt three models for the internal kinematics: isotropic
velocities, constant anisotropy and increasingly radial Osipkov-Merritt
anisotropy.
Analytical solutions are found for the radial
dependence of the mass, gravitational
potential, velocity dispersion, energy and virial ratio and we test their
variability with the concentration parameter describing the density
profile and amount of velocity anisotropy.
We also compute structural parameters, such as half-mass radius,
effective radius and various measures of concentration. Finally, we derive
projected quantities, the surface mass density and line-of-sight as well
as aperture velocity dispersion, all of which can be directly applied in
observational tests of current scenarios of structure formation.

On the mass scales of galaxies, if constant mass-to-light is assumed, the
NFW surface density profile is found to fit well Hubble-Reynolds
laws. It is also well fitted by S\'ersic $R^{1/m}$ laws, for $m \simeq 3$,
but in a much narrower range of $m$ and with much larger effective radii
than are observed. Assuming in turn reasonable values of the effective
radius, the mass density profiles imply a mass-to-light ratio that
increases outwards at all radii.

\end{abstract}


\section{Introduction}

A universal profile of dark matter haloes was introduced as a result
of high-resolution $N$-body simulations performed by Navarro, Frenk \&
White (1995, 1996, 1997, hereafter NFW) for power-law as well as CDM
initial power spectra of density fluctuations. NFW found that in a
large range of masses the density profiles of dark haloes can be fitted
with a simple formula with only one fitting parameter. The density profile
steepens from $r^{-1}$ near the centre of the halo to $r^{-3}$ at large
distances. The NFW profile has been confirmed in cosmological simulations
by Cole \& Lacey (1996), Tormen, Bouchet \& White (1997), Huss, Jain \&
Steinmetz (1999a), Jing (2000), Bullock et al. (1999), while Huss, Jain \&
Steinmetz (1999b) have shown that the NFW profile also arises from
non-cosmological initial conditions. It is worthwhile noting that some
(but not all) recent very high resolution cosmological simulations produce
steeper density profiles, with inner slopes $\simeq -1.5$ (Fukushige \&
Makino 1997, Moore et al. 1998, Ghigna et al. 1999, see also Jing \& Suto
2000). The density profiles in the cosmological simulations also display
considerable scatter (Avila-Reese et al. 1999, Bullock et al. 1999), and
Avila-Reese et al. find that the outer slopes of galaxy size haloes are
steeper than the NFW slope of $-3$ when selected within clusters ($-4$)
and slightly shallower within groups ($-2.7$). Although the exact
properties of dark matter haloes are still under debate, the NFW profile is
presently considered to provide the reference frame for any further
numerical research on density profiles of dark haloes.

Simple cosmological derivations of the density profiles of bound
objects are difficult, essentially because one needs to work in the
non-linear regime of the growth of gravitational instabilities.
Nevertheless, using the spherical top-hat model of Gunn \& Gott
(1972), density profiles typically varying as $r^{-9/4}$ were derived by
Gott (1975), Gunn (1977), Fillmore \& Goldreich (1984) and Bertschinger
(1985). Hoffman \& Shaham (1985) applied the spherical infall model to the
hierarchical clustering scenario and predicted that the density profiles of
haloes should depend on $\Omega$ as well as the initial power spectrum of
density fluctuations. However, for $\Omega=1$ they obtained power-law
profiles in contradiction with the steepening slopes found in the current
$N$-body simulations described above. In a recent study,  \L okas (2000)
has improved the model of Hoffman \& Shaham
(1985) by a generalization of the initial density distribution, the
introduction of a cut-off in this distribution at half the inter-peak
separation and by a proper calculation of the collapse factor. The
improved model reproduces the changing slope of the density profile and its
dependence on halo mass and the type of cosmological power spectrum found by
NFW. The NFW profile is also reproduced in studies taking into account
the merging mechanism (see Lacey \& Cole 1993) in the halo formation
scenario (e.g. Salvador-Sol\'{e}, Solanes \& Manrique 1998, Avila-Reese,
Firmani \& Hernandez 1998). Therefore the numerical and analytical
considerations seem to converge on the statement that the density
profiles of dark matter haloes are indeed well described by the universal
formula proposed by NFW.

The ultimate test of both the analytical and numerical results must come
from the observations of density profiles of galaxies and
galaxy clusters. Three recent studies of clusters (Carlberg et al. 1997,
Adami et al. 1998, van der Marel et al. 2000) claim good agreement between
cluster observations and the NFW mass density profile. But for galaxies,
the situation is less satisfying. Flores \& Primack (1994) show that the
NFW profile is incompatible with the rotation curves of spiral galaxies,
while Kravtsov et al. (1998) estimate that the inner slope of the density
profile of dwarf irregular and LSB galaxies is $-0.3$ instead of $-1$.
However, these conclusions were obtained with a number of assumptions and
approximations concerning the very unclear issues of biasing,
non-sphericity of objects and so on. Besides, as pointed out by van den
Bosch et al. (2000), Swaters, Madore \& Trewhella (2000) and van den Bosch
\& Swaters (2000), the observed rotation curves of these galaxies are too
uncertain to discriminate between cores and cusps.

The main motivation for this research is to explore analytically
the physical properties of objects with NFW density profiles.
The aim is to check whether these properties are acceptable from the
physical point of view and thus to test the validity of density profiles
obtained in cosmological $N$-body simulations. Additionally, this paper
presents formulae for observable quantities that can be used for
comparisons between the theoretical predictions (such as the NFW profile)
and observations.

The paper is organized as follows: after a short presentation of
the universal formula for the density profile proposed by NFW, in
Section~2 we describe physical properties of spherical systems following
from this density profile. Section~3 is devoted to a simple comparison between
the projected NFW density profile and the surface brightness of elliptical
galaxies. A more thorough comparison is beyond the scope of the present
paper and will be given elsewhere (Mamon \& {\L}okas, in preparation).
The discussion follows in Section~4.

\section{Properties of the NFW model}

\subsection{Basic properties}

NFW established that the density profiles of dark
matter haloes in high resolution cosmological simulations for a wide range
of masses and for different initial power spectra of density fluctuations
are well fitted by the formula
\begin{equation}    \label{c1}
    \frac{\rho(r)}{\rho_c^0} = \frac{\delta_{\rm char}}{(r/r_{\rm
    s})\,(1+r/r_{\rm s})^{2}}
\end{equation}
with a single fitting parameter $\delta_{\rm char}$, the characteristic
density. The so-called scale radius $r_{\rm s}$ is defined by
\begin{equation}     \label{c2}
    r_{\rm s} = \frac{r_{v}}{c} \ ,
\end{equation}
where $r_{v}$ is the virial radius usually defined as the distance from the
centre of the halo within which the mean density is $v$ times the
present critical density, $\rho_c^0$. The value of the virial
overdensity $v$ is often assumed to be $v=178$, a number predicted
by the simplest version of the spherical model for $\Omega=1$. For other
cosmological models it can be lower by a factor of 2 or more
(Lacey \& Cole 1993, Eke, Cole \& Frenk 1996). However, according
to the improved spherical infall model (\L okas 2000) $v$ can be as low
as 30 even for $\Omega=1$. In the following, $v$ is kept as a free parameter.

The quantity $c$ introduced in equation (\ref{c2})
is the concentration parameter, which is related to the
characteristic density by
\begin{equation}    \label{c3}
    \delta_{\rm char} = \frac{v \,c^3 g(c)}{3} \ ,
\end{equation}
where
\begin{equation}    \label{c4}
    g(c) = \frac{1}{\ln (1+c) - c/(1+c)} \ .
\end{equation}
The concentration parameter will be used hereafter as the only parameter
describing the shape of density profile. From cosmological
$N$-body simulations (Navarro et al. 1997, Jing 2000, Bullock et
al. 1999, Jing \& Suto 2000), extended Press-Schechter theory (Navarro et
al. 1997, Salvador-Sol\'e, Solanes \& Manrique 1998), and the spherical
infall model (\L okas 2000), we know that $c$ depends on the mass of
object and the form of the initial power spectrum of density fluctuations.
For all initial power spectra, the observed trend is for lower
concentration parameter in higher mass objects, with $4 < c < 22$ in
cosmological simulations with CDM initial power spectra and $c$ up to 90
for the less realistic scale-free power spectra. More precisely, in the
$\Lambda$CDM cosmology, $c=5$ corresponds to the masses of clusters of
galaxies, while $c=10$ corresponds to the masses of bright galaxies.

It is convenient to express the distance from the centre of the object in
units of the virial radius $r_{v}$:
\begin{equation}    \label{c5}
    s=\frac{r}{r_{v}}
\end{equation}
and the density profile of equation (\ref{c1}) then becomes
\begin{equation}    \label{c6}
     \frac{\rho(s)}{\rho_c^0} = \frac{v \,c^2 g(c)}{3 \,s\,(1+ c
     s)^2} \ .
\end{equation}

The mass of the halo is usually defined as the mass within the virial
radius:
\begin{equation}    \label{c7}
    M_v = \frac{4}{3} \,\pi \,r_{v}^3\, v \,\rho_c^0 \ .
\end{equation}
The distribution of mass in units of the virial mass follows from
equation (\ref{c6}):
\begin{equation}    \label{c8}
    \frac{M(s)}{M_v} = g(c) \left[ \ln (1+c s) - \frac{c s}{1 + c
    s}  \right]
\end{equation}
and we see that it diverges at large $s$, which is a disadvantage of the
model from a physical point of view.

The gravitational potential associated with the density distribution
(\ref{c6}) is
\begin{equation}    \label{c9}
    \frac{\Phi(s)}{V_v^2} = - g(c) \,\frac{\ln (1 + c s)}{s} \ ,
\end{equation}
where $V_v$ is the circular velocity at $r=r_v$:
\begin{equation}    \label{c10}
    V_v^2 = V^2(r_v) = \frac{G M_v}{r_v}
    = \frac{4}{3} \,\pi \,G \,r_{v}^2 \,v \,\rho_c^0 \ .
\end{equation}
Hence, from equation~(\ref{c9}) we see that
the gravitational potential at the centre, $\Phi(0) = - c g(c)
V_v^2$, is finite.

Equations (\ref{c8}) and (\ref{c10}) lead to a circular
velocity that obeys
\begin{equation}    \label{c12}
    \frac{V^2(s)}{V_v^2} = \frac{g(c)}{s} \left[ \ln (1+c s) -
    \frac{c s}{1 + c s}  \right] \ .
\end{equation}
Equations~(\ref{c8}), (\ref{c9}) and (\ref{c12}) were first derived by Cole
\& Lacey (1996).

The radial velocity dispersion $\sigma_{\rm r}(r)$ can be obtained by
solving the Jeans equation
\begin{equation}    \label{c13}
    \frac{1}{\rho} \frac{\rm d}{{\rm d} r} (\rho \sigma_{\rm r}^2) +
    2 \beta \frac{\sigma_{\rm r}^2}{r} = -\frac{{\rm d} \Phi}{{\rm d} r} \ ,
\end{equation}
where $\beta=1-\sigma_\theta^2(r)/\sigma_{\rm r}^2(r)$ is a measure of
the anisotropy in the velocity distribution. In the simplest case of
isotropic orbits, $\sigma_\theta(r)=\sigma_{\rm
r}(r)$ and $\beta=0$. This value of $\beta$ is also close to the results of
$N$-body simulations: Cole \& Lacey (1996) and Thomas et al. (1998) show
that, in a variety of cosmological models, the ratio
$\sigma_\theta/\sigma_{\rm r}$ is not far from unity and decreases slowly
with distance from the centre to reach $\simeq 0.8$ at the virial
radius. However, Huss, Jain \& Steinmetz (1999a) find
$\sigma_\theta/\sigma_{\rm r} \simeq 0.6$ at $r_v$.

First we consider the case of $\beta$=const.
Then the solution to the equation (\ref{c13}) with the condition of
$\sigma_{\rm r} \rightarrow 0$ at $s \rightarrow \infty$ is
\begin{eqnarray}
    \frac{\sigma_{\rm r}^2}{V_v^2} (s, \beta={\rm const})
    &=& g(c) (1+c s)^2 s^{1-2 \beta}
    \nonumber \\
    &\hspace{-2cm} \times & \hspace{-1.4cm} \int_{s}^\infty
    \left[ \frac{s^{2 \beta - 3} \ln (1+c
    s)}{(1+c s)^2} - \frac{c s^{2 \beta-2}}{(1+c s)^3} \right] {\rm d} s .
    \label{c13a}
\end{eqnarray}
For $\beta=0$, 0.5 and 1, reasonably simple analytical solutions to this
equation can be found:
\begin{eqnarray}
    \frac{\sigma_{\rm r}^2}{V_v^2} (s, \beta=0)
    &=& \frac{1}{2} c^2 \,g(c) \,s
    \,(1 + c s)^2 \ [\pi^2 - \ln (c s)  - \frac{1}{c s} \nonumber \\
    &\hspace{-2.8cm} - & \hspace{-1.7cm} \frac{1}{(1+c s)^2} -\frac{6}{1+c s}
    + \left(1 +\frac{1}{c^2 s^2} - \frac{4}{c s}
    - \frac{2}{1+c s} \right) \nonumber  \\
    &\hspace{-2.8cm} \times & \hspace{-1.7cm}  \ln (1+c s)
    + 3 \ln^2 (1+c s) + 6 \, {\rm Li}_2(-c s) ] \ ,
    \label{c14}
\end{eqnarray}
\begin{eqnarray}
    \frac{\sigma_{\rm r}^2}{V_v^2} (s, \beta=0.5)
    &=& c \, g(c) \, (1 + c s)^2
    [- \frac{\pi^2}{3} +\frac{1}{2(1+c s)^2}  \nonumber \\
    &\hspace{-2.8cm} + & \hspace{-1.7cm} \frac{2}{1+c s}
    +\frac{\ln (1+c s)}{c s} + \frac{\ln
    (1+c s)}{1+c s} \nonumber \\
    &\hspace{-2.8cm} - & \hspace{-1.7cm} \ln^2 (1+c s)
    - 2 {\rm Li}_2 (-c s)] , \label{c14a}
\end{eqnarray}
\begin{eqnarray}
    \frac{\sigma_{\rm r}^2}{V_v^2} (s, \beta=1)
    &=& g(c) \, (1 + c s)^2
    \frac{1}{s} \left[ \frac{\pi^2}{6} - \frac{1}{2(1+ c s)^2} \right.
    \nonumber \\
    &\hspace{-2.8cm} - & \hspace{-1.7cm} \left. \frac{1}{1+c s}
    - \frac{\ln (1+c s)}{1+c s} + \frac{\ln^2 (1+c s)}{2} +
    {\rm Li}_2 (-c s) \right].  \label{c14b}
\end{eqnarray}

In the above expressions ${\rm Li}_2 (x)$ is the dilogarithm, a special
function which can be conveniently dealt with using \emph{Mathematica}
packages. Otherwise, it can be approximated by
\begin{equation}   \label{dilog}
    {\rm Li}_2 (x) = \int_x^0 \!{\ln(1-t){\rm d}t \over t}
    \simeq x \left [1\!+\!10^{-0.5}(- x)^{0.62/0.7} \right ]^{-0.7} .
\end{equation}
The fit is accurate to better than 1.5\% in the range $-100 < x < 0$.

We included the predictions for $\beta=1$ just as a limiting case. In fact
such a model with purely radial orbits and NFW density profile is not
physical since its distribution function is not everywhere non-negative. As
pointed out by e.g. Richstone \& Tremaine (1984, see also \L okas \&
Hoffman 2000), such velocity anisotropy requires the inner density profile
to be $r^{-2}$ or steeper for the model to be physical.

A more realistic description of velocity anisotropy is provided by a model
proposed by Osipkov (1979) and Merritt (1985) with $\beta$ dependent on
distance from the centre of the object
\begin{equation} \label{c14c}
    \beta_{\rm OM} = \frac{s^2}{s^2 + s_{\rm a}^2}
\end{equation}
where $s_{\rm a}$ is the anisotropy radius determining the transition
from isotropic orbits inside to radial orbits outside.
As mentioned above, the results of $N$-body simulations suggest
$\sigma_{\theta}/\sigma_{\rm r} \simeq 0.8$ and therefore $\beta \simeq 0.36$
at $s=1$, which gives $s_{\rm a} \simeq 4/3$, a value that we adopt here
for all numerical calculations.

For the Osipkov-Merritt model the solution to the Jeans equation (with
the same boundary condition as before) reads
\begin{eqnarray}
    \frac{\sigma_{\rm r}^2}{V_v^2} (s, \beta_{\rm OM})
    &=&  \frac{g(c) s (1+c s)^2}{s^2 + s_{\rm a}^2}    \nonumber \\
    &\hspace{-2.5cm} \times & \hspace{-1.5cm} \int_{s}^\infty
    \left[ \frac{(s^2 + s_{\rm a}^2) \ln (1+c s)}{s^3 (1+c s)^2}
    - \frac{c (s^2 + s_{\rm a}^2)}{s^2 (1+c s)^3} \right] {\rm d} s
    \label{c14d}
\end{eqnarray}
and the integration gives
\begin{eqnarray}
    \frac{\sigma_{\rm r}^2}{V_v^2} (s, \beta_{\rm OM})
    &=& \frac{g(c) s (1+c s)^2}{2 (s^2 + s_{\rm a}^2)}
    \nonumber \\
    &\hspace{-3.5cm} \times & \hspace{-2cm}
    \left\{ -\frac{c s_{\rm a}^2}{s} - c^2 s_{\rm a}^2 \ln (c s)
    +  c^2 s_{\rm a}^2 \ln (1+c s)
    \left( 1 +\frac{1}{c^2 s^2} - \frac{4}{c s} \right) \right. \nonumber \\
    &\hspace{-3.5cm} - & \hspace{-2cm} (1+ c^2 s_{\rm a}^2)
    \left[ \frac{1}{(1+c s)^2} + \frac{2 \ln (1+c s)}{1+c s} \right]
    \label{c14e} \\
    &\hspace{-3.5cm} + & \hspace{-2cm} \left. (1+3 c^2 s_{\rm a}^2)
    \left[ \frac{\pi^2}{3} - \frac{2}{1+c s} + \ln^2 (1+c s)
    + 2 {\rm Li}_2(-c s) \right] \right\} . \nonumber
\end{eqnarray}

\begin{figure}
\begin{center}
    \leavevmode
    \epsfxsize=8cm
    \epsfbox[50 50 340 560]{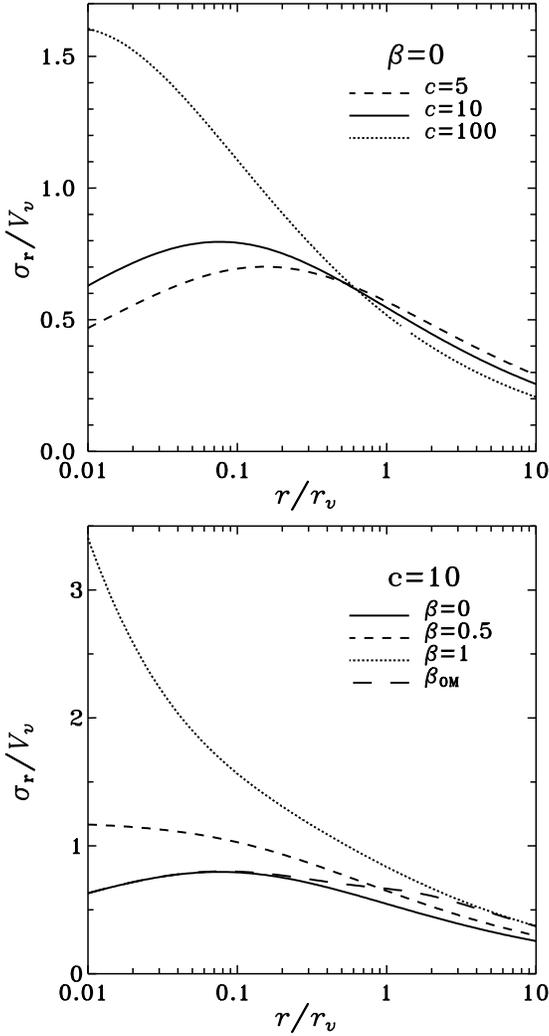}
\end{center}
    \caption{Radial velocity dispersion profile (in units of the
    circular velocity at the virial radius), given by in isotropic model,
    equation (\ref{c14}), for three different values of the concentration
    parameter $c$ (upper panel) and for the four considered anisotropy
    models with $c=10$ (lower panel).}
\label{sig}
\end{figure}

Figure~\ref{sig} shows the radial dependence of the radial velocity
dispersion. The upper panel of the Figure presents how the results depend on
the concentration parameter in the isotropic case, while the lower panel
compares predictions for different anisotropy models with $c=10$.

\subsection{The energy distributions}

\begin{figure}
\begin{center}
    \leavevmode
    \epsfxsize=8cm
    \epsfbox[50 50 340 560]{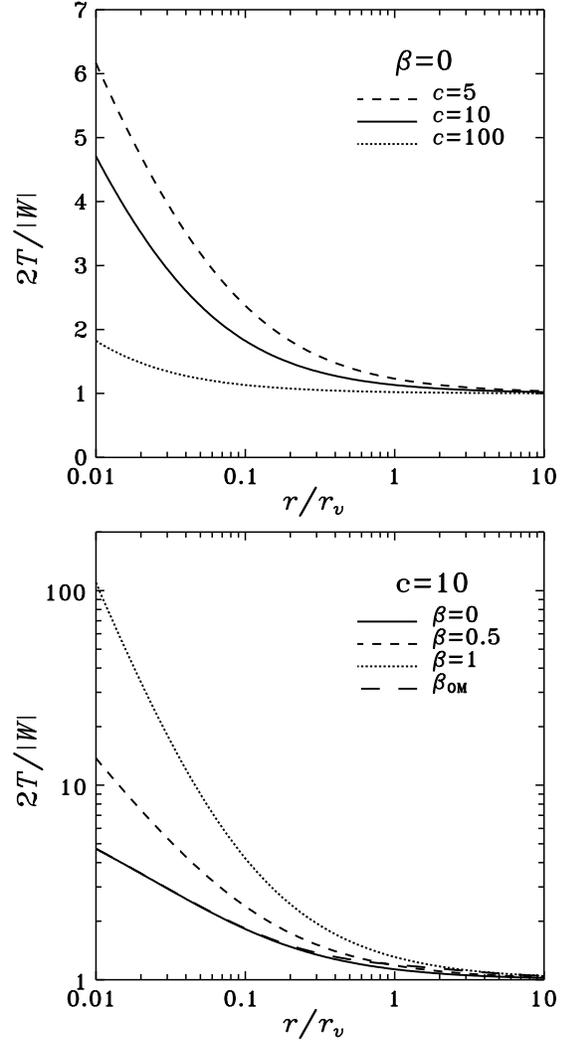}
\end{center}
    \caption{The radial dependence
    of the virial ratio in the isotropic model (eqs.
    [\ref{c15}] and [\ref{c17}]) for three different values of the
    concentration parameter (upper panel) and for the four considered
    anisotropy models with $c=10$ (lower panel).}
\label{tws}
\end{figure}

\begin{figure}
\begin{center}
    \leavevmode
    \epsfxsize=8cm
    \epsfbox[50 50 340 310]{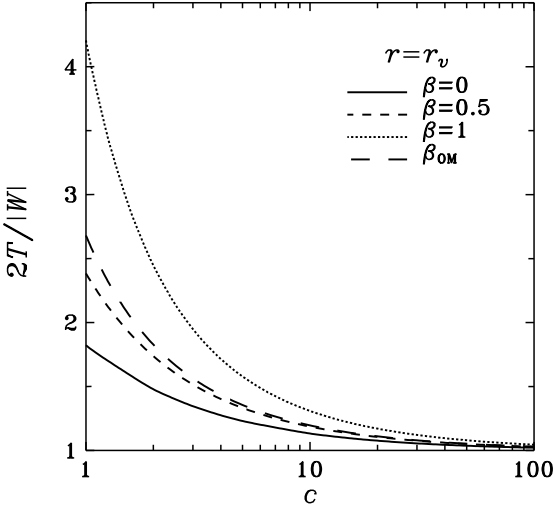}
\end{center}
    \caption{Dependence on the concentration parameter
    of the virial ratio at the virial radius for the four considered
    anisotropy models.}
\label{twc}
\end{figure}

The potential energy associated with the mass distribution of equation
(\ref{c8}) is
{\samepage
\begin{eqnarray}
    W(s) &=& - \frac{1}{r_v} \int_0^s \frac{G M(s)}{s}
    \frac{{\rm d} M(s)}{{\rm d} s} {\rm d} s \nonumber \\
    &=& - W_\infty \left[ 1-\frac{1}{(1+c s)^2}
    - \frac{2 \ln (1+c s)}{1+c s} \right] \ , \label{c15}
\end{eqnarray}
}
where
\begin{equation}    \label{c16}
    W_\infty = - \lim_{s \rightarrow \infty} W(s) =
    \frac{c g^2(c) G M_v^2}{2 r_v} .
\end{equation}
The kinetic energy for arbitrary $\beta$ is given by
\begin{equation}    \label{c16a}
    T(s, \beta) = 2  \pi \,r_v^3
    \,\int_0^s (3 - 2 \beta) \rho (s) \sigma_{\rm
    r}^2(s, \beta) \,s^2\, {\rm d} s .
\end{equation}
For the three cases of $\beta$=0, 0.5 and 1, we obtain respectively
\begin{eqnarray}
    T(s, \beta=0) &=&
    \frac{1}{2} W_\infty \{ -3 + \frac{3}{1+c s} - 2 \ln (1+c s)
    \nonumber \\
    &\hspace{-2.8cm} + & \hspace{-1.7cm} c s [5 + 3 \ln (1+ c s)]
    - c^2 s^2 [7 + 6 \ln (1+ c s)]
    \nonumber \\
    &\hspace{-2.8cm} + & \hspace{-1.7cm} c^3 s^3 [\pi^2 - \ln c - \ln s +
    \ln (1+c s) \nonumber \\
    &\hspace{-2.8cm} + & \hspace{-1.7cm} 3 \ln^2 (1+c s) +
    6 {\rm Li}_2 (-c s)] \} \ ,
    \label{c17}
\end{eqnarray}
\begin{eqnarray}
    T(s, \beta=0.5) &=&
    \frac{1}{3} W_\infty \{ -3 + \frac{3}{1+c s} - 3 \ln (1+c s)
    \nonumber \\
    &\hspace{-2.8cm} + & \hspace{-1.7cm}
    6 c s [1 + \ln (1+ c s)] - c^2 s^2 [\pi^2
    \nonumber \\
    &\hspace{-2.8cm} + & \hspace{-1.7cm} 3 \ln^2 (1+c s)
    + 6 {\rm Li}_2 (-c s)] \} \ ,
    \label{c17a}
\end{eqnarray}
\begin{eqnarray}
    T(s, \beta=1) &=&
    \frac{1}{2} W_\infty \{ - 2 \ln (1+c s)
    + c s [\frac{\pi^2}{3} - \frac{1}{1+ c s}
    \nonumber \\
    &\hspace{-2.8cm} + & \hspace{-1.7cm} \ln^2 (1+c s)
    + 2 {\rm Li}_2 (-c s)] \} ,
    \label{c17b}
\end{eqnarray}
where we have used in each case the corresponding expression for
$\sigma_{\rm r}^2(s, \beta)$ from equations (\ref{c14})-(\ref{c14b}). For the
Osipkov-Merritt model the calculation has to be done numerically.

The results for the potential and kinetic energy (\ref{c15})-(\ref{c17b})
lead to a virial ratio $\lim_{s \rightarrow \infty} 2 T/|W| = 1$
for any value of $c$, in agreement with the virial theorem.
Figure~\ref{tws} shows how the virial ratio depends on distance for three
different values of the concentration parameter in the isotropic case (upper
panel) and compares the ratios obtained for different $\beta$ with $c=10$
(lower panel). At low radii, the virial ratio is large, especially for low
concentration parameters and models with much anisotropy. However, as
demonstrated by Figure~\ref{twc}, at the virial radius $r_v (s\!=\!1)$,
$2T/|W|$ is still greater than unity and grows with the amount of
anisotropy in the model. We see that the virial theorem is better
satisfied at $s=1$ for objects with larger concentration parameters, as
$\lim_{c \rightarrow \infty} 2 T/|W| (s=1) = 1$.
Since objects of smaller mass have larger concentration parameters, they
are closer to dynamical equilibrium.

The scalar virial theorem we referred to above is expected to be satisfied
for self-gravitating systems in steady state. In more realistic
situations, the system is never isolated and experiences an external
gravitational field; there is also continuous infall of matter. We may
conclude from the results above that objects with NFW density
profiles and different velocity distributions are close to dynamical
equilibrium. However, the virial ratio cannot be used to define the
boundary of the virialized object.

\subsection{Structural parameters}

\begin{figure}
\begin{center}
    \leavevmode
    \epsfxsize=8cm
    \epsfbox[50 50 340 310]{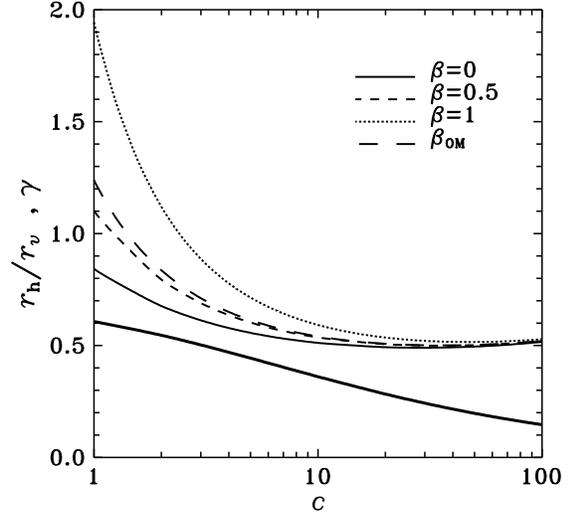}
\end{center}
\caption{Dependence on the concentration parameter of the half-mass
radius, scaled to the virial radius (thicker lower solid line, see
eq.~[\ref{rhfit}]) and $\gamma$ for the four anisotropy models
(eq.~[\ref{gammadef2}], four upper lines).}
\label{rhgamma}
\end{figure}

A useful quantity is the half-mass radius. Unfortunately, the
divergence of the mass of the NFW profile forces one to define the
half-mass radius within a cutoff radius $r_{\rm cut}$. The most natural
choice is $r_{\rm cut} = r_v$, since the density distribution is only
reliable out to the virial radius. With $r_{\rm cut} = r_v$, the
half-mass radius $r_{\rm h}$ satisfies the following relation for
the mass of dimensionless radius:
\begin{equation}
    M\left({r_{\rm h}\over r_v}\right) = {M(1)\over 2} \ .
\end{equation}
Numerical values of $r_{\rm h}/r_v$ are easily obtained using
equation (\ref{c8}) and over the range $1 < c < 100$ they can be
approximated to better than 2\% accuracy by
\begin{eqnarray}    \label{rhfit}
    {r_{\rm h}  \over r_v} &=& 0.6082 - 0.1843\,\log c \nonumber \\
    & & \mbox{} - 0.1011\,\log^2 c
    + 0.03918\,\log^3 c \ .
\end{eqnarray}
The lowest thick solid line in Figure~\ref{rhgamma} shows how
$r_{\rm h}/r_v$ decreases with increasing concentration parameter.

It is useful to estimate the concentration $\gamma$ of a dynamical system,
such that
\begin{equation}    \label{gammadef}
    \left\langle \sigma^2 \right \rangle = \gamma {G M \over r_{\rm h}} \ ,
\end{equation}
where $\langle \sigma^2 \rangle =\langle \sigma_{\rm r}^2 + \sigma_{\theta}^2
+ \sigma_{\phi}^2 \rangle $ is the mass weighted mean-square velocity
dispersion. As first noted by Spitzer (1969) for polytropes, many
realistic density profiles have $\gamma = 0.4$. For example, it is easy
to show that for the Hernquist (1990) model with $\beta=0$, $\gamma
= (1+\sqrt{2})/6 \simeq 0.403$ (Mamon 2000).

Using equation~(\ref{gammadef}) and limiting again the mass to $r_{\rm cut} =
r_v$, we define $\gamma$ with
\begin{equation}    \label{gammadef2}
    \gamma = {r_{\rm h}\,\left \langle \sigma^2
    \right\rangle_{r \leq r_v} \over G M(1)} = 2\,{r_{\rm h}\,T(1, \beta)
    \over G M^2(1)} \ ,
\end{equation}
where we made use of
\begin{equation}    \label{tvssigmav}
    T(x, \beta) = {1 \over 2}\,M(x)\,\left\langle \sigma^2
    \right\rangle_{r \leq x\,r_v} \ .
\end{equation}
The values of $\gamma$ for different velocity anisotropy models, derived
from equations~(\ref{c7}), (\ref{c8}), (\ref{c16}), (\ref{c16})-(\ref{c17b}),
(\ref{rhfit}), and (\ref{gammadef2}) are shown in Figure~\ref{rhgamma}
and in the case of $\beta=0$ yield numbers closest to 0.4: $\gamma = 0.56$
for $c = 5$ and $\gamma = 0.51$ for $c = 10$. Thus the NFW model produces
$\gamma$s that are higher than the canonical value of 0.4, especially if more
velocity anisotropy is assumed. This may be caused by the ill-defined
cutoff radius.

In models with homogeneous cores, the central density, the core radius
$r_{\rm c}$ and the central 3-D velocity dispersion $\sigma^2 (0)$ are
related through
\begin{equation}
    4\pi G\rho(0) r_{\rm c}^2 = \frac{1}{3} \eta \,\sigma^2 (0) \ .
\end{equation}
King (1966) models have $\eta = 9$.
In models with cuspy cores, we propose the scaling relation
\begin{equation}
    4\pi G\rho(r_{\rm s}) r_{\rm s}^2 =
    \frac{1}{3} \eta \left \langle \sigma^2 \right\rangle_{r < r_{\rm s}} \ .
\end{equation}
Using equations~(\ref{c2}), (\ref{c6}) and (\ref{c7}), one has
$4 \pi G \rho(r_{\rm s}) r_{\rm s}^2 = c\,g(c)\,V_v^2/4$ and from
equation~(\ref{tvssigmav}) for $x = 1/c$
one obtains
\begin{equation}
    \eta = \frac{3 c g(c) V_v^2 M(1/c)}{8 T(1/c, \beta)} \ .
\end{equation}
For different velocity anisotropy models we then have
\begin{equation}
    \eta (\beta=0) = \frac{3(2 \ln 2 - 1)}{2(\pi^2 - 7 - 8 \ln 2
    + 6 \ln^2 2)} \simeq 2.797 ,
\end{equation}
\begin{equation}
    \eta (\beta=0.5) = \frac{9(1 - 2 \ln 2)}{4(\pi^2 - 9 - 6 \ln 2
    + 6 \ln^2 2)} \simeq 2.138 ,
\end{equation}
\begin{equation}
    \eta (\beta=1) = \frac{9(2 \ln 2 - 1)}{2(\pi^2 - 3 - 12 \ln 2
    + 6 \ln^2 2)} \simeq 1.212 ,
\end{equation}
where we have used equations~(\ref{c8}) and (\ref{c17})-(\ref{c17b}), and
the fact that ${\rm Li}_2(-1) = -\pi^2/12$. Note that $\eta$ is independent
of $c$ in all cases with $\beta=$const. For the Osipkov-Merritt model $\eta$
is no longer a constant but we find $1.902 < \eta < 2.797$ in the range
$1 < c < 100$ with the limiting cases of $\eta \rightarrow \eta(\beta=1)$ for
$c \rightarrow 0$ and $\eta \rightarrow \eta(\beta=0)$ for
$c \rightarrow \infty$. Such limiting behaviour is due to the fact that
for large $c$ the integration of $T(1/c, \beta)$, equation (\ref{c16a}),
probes only the range of $s$ where $\beta$ is close to zero, while for
small $c$ the integral is dominated by contribution from large $s$ where
$\beta$ is close to unity.

Finally, we consider the structural parameter
\begin{equation}
    \hbox{WUM} = {W(s) \over M(s) \Phi(0)}
\end{equation}
brought forward by Seidov \& Skvirsky (2000) with the motivation of WUM
being constant for different self-gravitating objects of simple geometry.
Using equations (\ref{c8}), (\ref{c9}) and (\ref{c15}) we find that for
the NFW model
\begin{equation}
    \hbox{WUM} = \frac{ c s (2 + c s) - 2 (1+ c s) \ln (1+c s)}{2 (1+ c s)
    [- c s + (1+ c s) \ln (1+ c s)]}
\end{equation}
so the parameter turns out to be a function of $c s = r/r_{\rm s}$ only.
It grows with $s$ from zero at $s \rightarrow 0$ reaching a maximum
value of $0.196$ at $r/r_{\rm s}=4.62$ and decreases
to zero again as $s \rightarrow \infty$. The values of this parameter at
the virial radius are $0.196$, $0.187$ and $0.125$ respectively for $c=5,
10$ and $100$.

\subsection{The distribution function}

\begin{figure}
\begin{center}
    \leavevmode
    \epsfxsize=8cm
    \epsfbox[50 50 340 310]{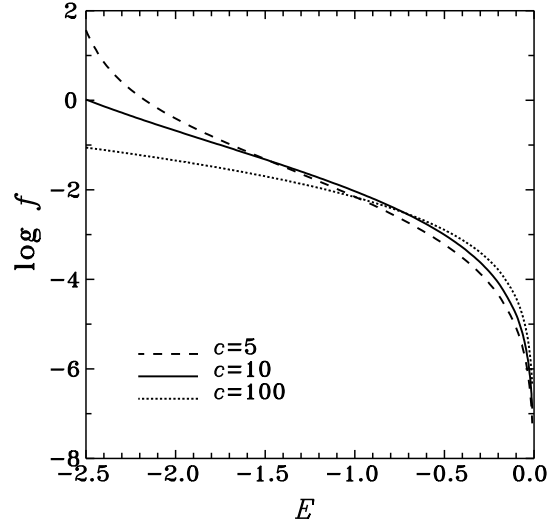}
\end{center}
    \caption{The distribution function for isotropic model (eq. [\ref{c26}])
    for three different values of the concentration parameter.}
\label{df}
\end{figure}

A quantity of great dynamical importance is the distribution function. For
a spherical system with an isotropic velocity tensor, the distribution
function depends on the phase-space coordinates only through the energy
(e.g. Binney \& Tremaine 1987), and can be derived through the Eddington
(1916) formula (e.g. Binney \& Tremaine 1987):
\begin{equation}     \label{c26}
    f({\cal E}) = \frac{1}{\sqrt{8} \pi^2} \left [\int_{0}^{{\cal E}}
    \frac{{\rm d}^2 \rho}{{\rm d} \Psi^2} \frac{{\rm d}
    \Psi}{\sqrt{{\cal E} - \Psi}} + \frac{1}{{\cal E}^{1/2}}
    \left ( \frac{{\rm d} \rho}{{\rm d}\Psi}\right)_{\Psi=0}\right ] \ ,
\end{equation}
where ${\cal E}$ and $\Psi$ are the conventionally defined relative
energy and potential; here ${\cal E} = -E$, where $E$ is the total
energy per unit mass and $\Psi = -\Phi$, where $\Phi$ is given by
equation (\ref{c9}).

It is easy to show that, given equations (\ref{c6}) and (\ref{c9}), the
second term in brackets in equation~(\ref{c26}) is zero.
The simplest way to perform the integration of the first term is to
introduce dimensionless variables
$\widetilde{\Psi} = \Psi/C_1$ and $\widetilde{\rho} = \rho/C_2$,
where $C_1 = g(c)\,V_v^2$ and $C_2 =  c^2 g(c) M_v/(4 \pi r_v^3)$. Then
the integration variable should be changed to $s$ and the limit of integration
corresponding to $\cal E$ found numerically for each $\cal E$ by solving
equation $\Psi(s)=\cal E$. Otherwise, with a few percent accuracy, the
integration in (\ref{c26}) can be done directly with an approximation
$s_{\rm apx } = -1.75 \ln (\widetilde{\Psi}/c)/\widetilde{\Psi}$.

The calculations of the distribution function are usually performed in
units such that $G=M=R_{\rm e}=1$ (Binney \& Tremaine 1987), where $M$ is
the total mass of the system and $R_{\rm e}$ is its effective radius.
Since in the case of NFW profile the total mass is infinite a reasonable
choice seems to be to put $M_v = 1$. The effective radius is not well
defined either but can be approximated as $r_v/2$ (see the next
section). Therefore we choose the units so that $G = M_v = r_v/2 = 1$
and arrive at the numerical results shown in Figure~\ref{df}. This choice
of normalization is equivalent to measuring $f$ in units of $\sqrt{8} M_v
/(r_v V_v)^3$ and $E$ in units of $V_v^2$.

Figure~\ref{df} proves that the distribution function turns out to be
similar to the distribution functions obtained from other density
profiles (see e.g. Figure 4-12 in Binney \& Tremaine 1987), except that
the NFW distribution functions do not display the cutoff at nearly unbound
energies characteristic of King (1966) models. The results
shown in Figure~\ref{df} indicate a proper behaviour of the
distribution function (it is nowhere negative). Quantitative comparisons
with other models should, however, be made with caution because of the
aforementioned problem with normalization. Distribution functions for
more realistic velocity dispersion models, like the Osipkov-Merritt model,
were recently considered in detail by Widrow (2000).

\subsection{Projected distributions}

\begin{figure}
\begin{center}
    \leavevmode
    \epsfxsize=8cm
    \epsfbox[50 50 340 560]{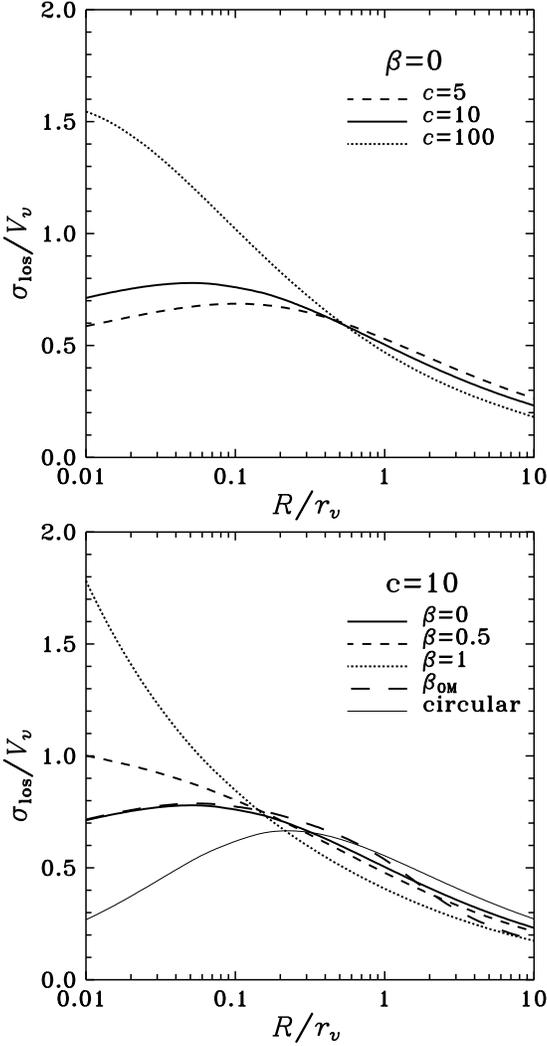}
\end{center}
    \caption{Upper panel: radial dependence of the line-of-sight
    velocity dispersion for isotropic orbits (eq. [\ref{c24}]) on the
    projected radius for three values of the concentration parameter.
    Lower panel:
    comparison of the line-of-sight velocity dispersion profiles for
    four anisotropy models calculated with $c=10$.}
\label{losvd}
\end{figure}

\begin{figure}
\begin{center}
    \leavevmode
    \epsfxsize=8cm
    \epsfbox[50 50 340 560]{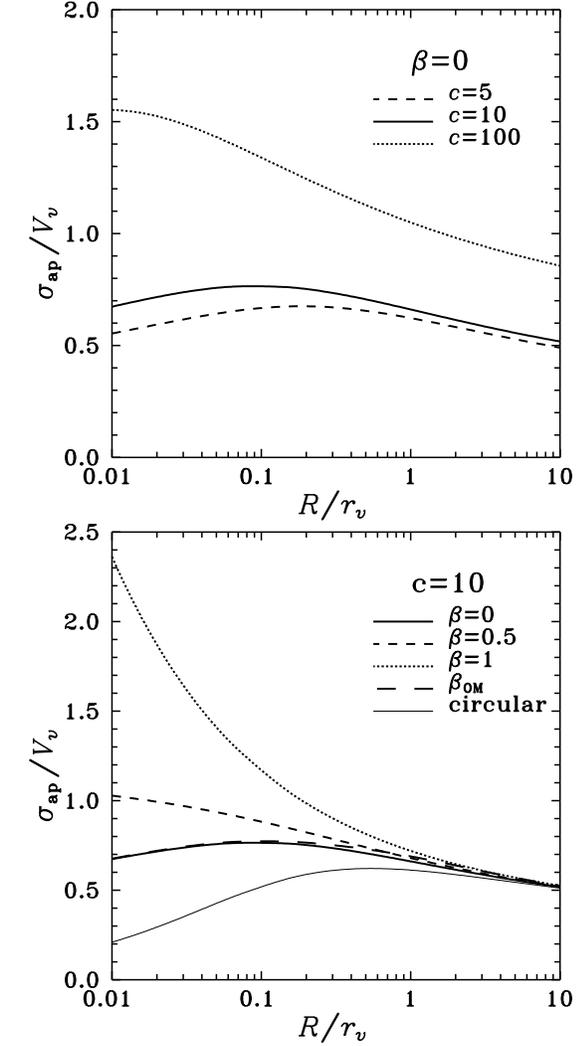}
\end{center}
    \caption{Upper panel: radial profiles of the aperture velocity
    dispersion in the isotropic model for three concentration parameters.
    Lower panel:
    comparison of the aperture velocity dispersions for four anisotropy
    models calculated with $c=10$.}
\label{sap}
\end{figure}

Of primary importance for comparisons with observations are the
projected distributions. The surface mass density of an object is
obtained by integrating the density along the line of sight:
\begin{eqnarray}
    \Sigma_M(R) \!\!\!\!&=& \!\!\!\!
    2\, \int_{R}^{\infty}
    \frac{r \,\rho(r)}{(r^2 - R^2)^{1/2}} \,{\rm d} r \nonumber \\
    &=& \!\!\!\!\frac{c^2\,g(c)}{2\pi}\,\frac{M_v}{r_v^2}\,
    \frac{1 - |c^2{\widetilde R}^2 - 1|^{-1/2}  C^{-1} [1/(c\widetilde
    R)]}{(c^2{\widetilde R}^2 -1)^2}  ,
\label{c20}
\end{eqnarray}
where
\begin{equation}      \label{c20a}
    C^{-1} (x) = \left\{
    \begin{array}{ll}
    \cos^{-1} (x) & \mbox{if $R > r_{\rm s}$} \\
    \cosh^{-1} (x)  & \mbox{if $R < r_{\rm s}$ .}
    \end{array}
    \right.
\end{equation}
In the above expressions $R$ is the projected radius and ${\widetilde
R}=R/r_v$. For the singular case $R = r_{\rm s}$ the $\widetilde
R$-dependent expression in equation (\ref{c20}) equals $1/3$ and we have
$\Sigma_M(R) = c^2 g(c) M_v/(6 \pi r_v^2)$ .
An analytical formula equivalent to equation (\ref{c20}) was derived
independently by Bartelmann (1996).

The projected mass is then given by
\begin{eqnarray}
    M_{\rm p}(R) \!\!\!\!&=&\!\!\!\!
    2 \pi \int_{0}^{R} R\,\Sigma_M(R) \,{\rm d} R \nonumber  \\
    &=& \!\!\!\! g(c) \,M_v
    \left[\frac{C^{-1} [1/(c\widetilde R)]}
    {|c^2{\widetilde R}^2 -1|^{1/2}} + \ln
    \left ( \frac{c\widetilde R}{2}\right)\right] \ , \label{c21}
\end{eqnarray}
which is logarithmically divergent at large ${\widetilde R}$. $C^{-1}(x)$
is again given by equation (\ref{c20a}).

Another important projected quantity is the line-of-sight velocity
dispersion which for a spherical non-rotating system
is (Binney \& Mamon 1982)
\begin{equation}    \label{c24}
    \sigma_{\rm los}^2 (R) = \frac{2}{\Sigma_M(R)} \int_{R}^{\infty}
    \left( 1-\beta \frac{R^2}{r^2} \right) \frac{\rho \,
    \sigma_{\rm r}^2 (r, \beta) \,r}{\sqrt{r^2 - R^2}} \,{\rm d} r \ ,
\end{equation}
where $\Sigma_M(R)$ is given by equation (\ref{c20}) and the radial
velocity dispersions $\sigma_{\rm r}(r, \beta)$ for our four models are
given by equations (\ref{c14})-(\ref{c14b}) and (\ref{c14e}).
For circular orbits, $\sigma_{\rm r}=0$, and one has
\begin{equation}    \label{c25}
    \sigma_{\rm los}^2 (R) = \frac{1}{\Sigma_M(R)} \int_{R}^{\infty}
    \left( \frac{R}{r} \right)^2 \frac{\rho \,V^2 \,r}{\sqrt{r^2 -
    R^2}}\, {\rm d} r \ ,
\end{equation}
where $V$ is the circular velocity given by equations (\ref{c10}) and
(\ref{c12}). The upper panel of Figure~\ref{losvd} shows the profiles of
line-of-sight velocity dispersion (with isotropic orbits), obtained
through numerical integration of equation (\ref{c24}) for different
concentration parameters. The lower panel of Figure~\ref{losvd} compares
the radial profiles of line-of-sight velocity dispersions obtained
for $c=10$ for different velocity anisotropy models.

For more distant or intrinsically small galaxies, as well as for groups
and clusters, spectroscopic observations are often limited to a single
large aperture centred on the object.
The mean velocity dispersion within an aperture (hereafter,
aperture velocity dispersion) is
\begin{equation}     \label{c28}
    \sigma_{\rm ap}^2 (R) = \frac{S^2(R)}{M_{\rm p}(R)}
\end{equation}
where
\begin{equation}    \label{c29}
   S^2(R) = 2 \pi \int_0^R \Sigma_M(P) \sigma_{\rm los}^2 (P) P {\rm d} P.
\end{equation}
In the above expressions $R$ is the radius of the aperture,
$\Sigma_M(P)$ is the surface mass distribution, equation (\ref{c20}),
and $M_{\rm p}(R)$ is the projected mass given by equation (\ref{c21}).

Inserting the expression for $\sigma_{\rm los}$ (eq.~[\ref{c24}])
into equation (\ref{c29}), we obtain a double integral, which after inversion
of the order of integration is reduced to an easily computable single
integral:
\begin{eqnarray}
    S^2(R) &=& c^2 \,g(c) \,M_v \left\{ \int_0^\infty
    \frac{\sigma_{\rm r}^2(s, \beta)\, s}{(1+c
    s)^2} \, \left(1-\frac{2 \beta}{3} \right) {\rm d} s \right.
    \nonumber \\
    &\hspace{-2cm} + & \hspace{-1.4cm}
    \left. \int_{\widetilde R}^\infty \frac{\sigma_{\rm
    r}^2(s, \beta) \,({s^2 - {\widetilde R}^2})^{1/2}}{(1+c s)^2} \left[
    \frac{\beta({\widetilde R}^2 + 2 s^2)}{3 s^2} - 1 \right] {\rm d} s
    \right\} , \label{c30}
\end{eqnarray}
where as before, $\widetilde R=R/r_v$, $s=r/r_v$ and $\sigma_{\rm
r}^2 (s, \beta)$ for different $\beta$ are given by
equations~(\ref{c14})-(\ref{c14b}) and (\ref{c14e}).
Analogous expression for circular orbits can be obtained from (\ref{c30})
by replacing $\sigma_{\rm r}^2$ by $V^2$, keeping only the terms
proportional to $\beta$ and dividing by $(-2 \beta)$.

Figure~\ref{sap} displays the radial profiles of aperture velocity
dispersion, computed numerically from equation~(\ref{c30}). From the upper
panel of the Figure we see that in the isotropic case the dependence of the
results on the concentration parameter is rather strong and monotonic for
a given $R$. The lower panel of the Figure compares the predictions for
different velocity anisotropy models.

\section{Comparison with observations}

\begin{figure}
\begin{center}
    \leavevmode
    \epsfxsize=8cm
    \epsfbox[90 40 550 779]{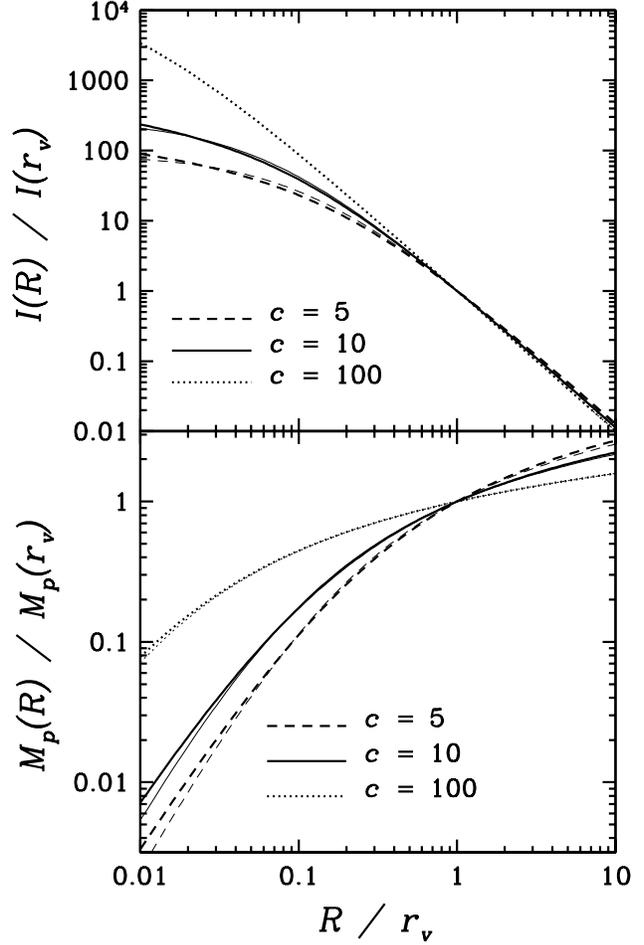}
\end{center}
    \caption{Radial profiles of the surface mass density,
    given by equation (\ref{c20}), (upper panel) and the
    projected mass, equation (\ref{c21}), (lower panel)
    for three different values of the concentration parameter.
    Hubble-Reynolds fits from equation~(\ref{c32}) are shown as
    thin curves ($R_{\rm HR}/r_v = 0.119$, $0.0640$ and
    $0.00743$ for $c=5, 10$ and $100$, respectively).
    For $c=100$, the NFW surface mass density is virtually
    indistinguishable from the best-fitting Hubble-Reynolds law.}
\label{is}
\end{figure}

\begin{figure}
\begin{center}
    \leavevmode
    \epsfxsize=8cm
    \epsfbox[50 50 340 310]{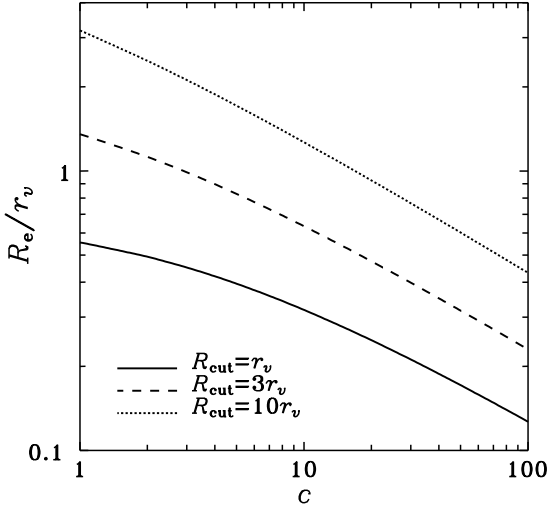}
\end{center}
    \caption{The dependence of the effective radius, defined in equation
    (\ref{c22}),  on the concentration parameter, with various choices of
    ${\widetilde R}_{\rm cut}$.}
\label{aec}
\end{figure}

Comparisons of the surface mass density to surface brightness
observations are usually performed with the assumption of constant
mass-to-light ratio $\Upsilon = {\rm const}$.
This assumption is not likely to be physical, because of the different
physics involved in the assemblies of the dark matter and baryonic components
of galaxies. In particular, the baryons in elliptical galaxies may well
settle at an early epoch, within a
radius that is the lower of the radius with virial overdensity $v \simeq
200$ and the radius at which gas can cool to form molecular clouds and later
stars. The baryons in ellipticals will then sit today in a region of
overdensity $v \gg 200$, and one then expects $\Upsilon$ to rise with $r$,
at least at large radii.

Nevertheless, for simplicity, we check whether the observations of elliptical
galaxies are consistent with the idea that stars are distributed
within elliptical galaxies according to the NFW density profile,
characterized by a virial radius where the mean overdensity is 200.
Such a situation may arise if the dark matter were negligible within
elliptical galaxies or distributed precisely like the luminous matter.
In a forthcoming paper (Mamon \& {\L}okas, in preparation), we will check
in more detail whether the observations of elliptical galaxies are
compatible or not with NFW density profiles for the mass distribution.

For constant mass-to-light ratio we have $\Sigma_M(R) =
\Upsilon I(R)$, where $I$ is the surface brightness. The radial profiles of
$I = \Sigma_M/\Upsilon$ and $M_{\rm p}$ are shown in Figure~\ref{is}. Both
quantities are normalized to their values at the virial radius.
Figure~\ref{is} shows that the surface mass density depends weakly on the
concentration parameter, especially at larger distances from the centre.

Since the surface mass density (eq. [\ref{c20}]) behaves as $1/R^2$ at
large distances, one may therefore compare it with the Hubble-Reynolds
formula (Reynolds 1913), which was the first model used to describe the
surface brightness profiles of elliptical galaxies:
\begin{equation}     \label{c32}
    I_{\rm HR} (R) = \frac{I_0}{(1+R/R_{\rm HR})^2} \ .
\end{equation}
$R_{\rm HR}$ is the characteristic radius of the distribution,
where the surface brightness falls to one-quarter of its central value.
The thin curves of Figure~\ref{is} show that the surface mass density of
the NFW model (eq.~[\ref{c20}]) is very well fitted by
equation~(\ref{c32}) and the best-fit values of ${\widetilde R}_{\rm HR}
= R_{\rm HR}/r_v$ are $0.119$, $0.0640$ and $0.00743$ respectively for
$c=5, 10$ and $100$.

\begin{figure}
\begin{center}
    \leavevmode
    \epsfxsize=9cm
    \epsfbox{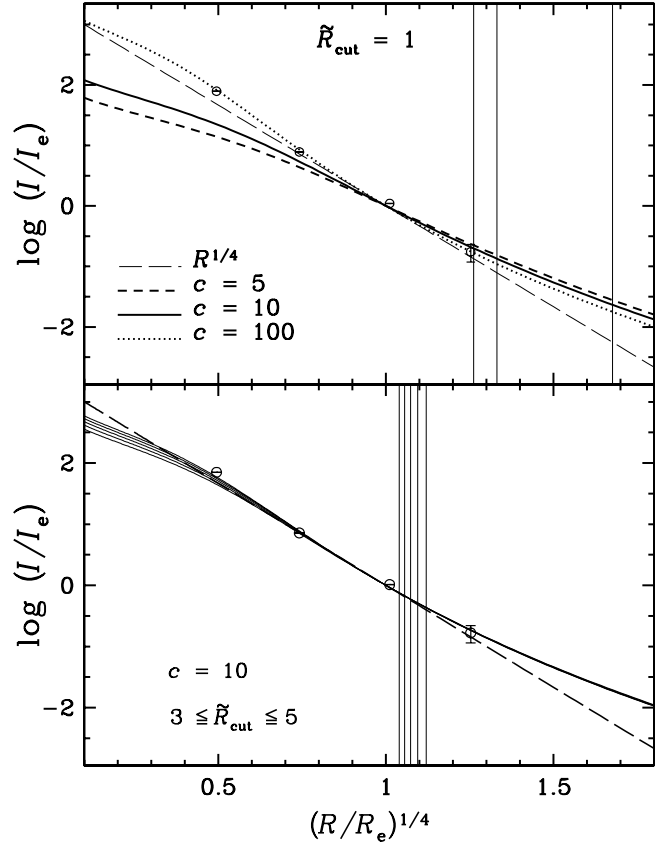}
\end{center}
    \caption{Upper panel: surface brightness profiles (eq.
    [\ref{c20}]) for three concentration parameters and ${\widetilde
    R}_{\rm cut}=1$. Lower panel: the dependence of the surface
    brightness profiles on the cut-off $\widetilde R_{\rm cut}$ for $c=10$
    and ${\widetilde R}_{\rm cut} = 3$, 3.5, 4, 4.5, and 5 (bottom
    to top curves). In both panels, the $R^{1/4}$ law (eq. [\ref{c23}])
    is shown as long dashed lines. The vertical lines
    represent the virial radius (for the three concentration parameters in
    the upper panel and for the 5 values of $\widetilde R_{\rm cut}$ in
    the lower panel, with $\widetilde R_{\rm cut}$ increasing from right
    to left). The circles are the data points for the galaxy NGC 3379.}
\label{de}
\end{figure}

The surface brightness profiles of astrophysical objects are often scaled
with the effective radius, which we denote $R_{\rm e}$,
where the projected luminosity is half the total luminosity. Given the
divergence of the projected mass, we are forced again to introduce a cut-off
at some scale $R_{\rm cut} = {\widetilde R}_{\rm cut}\,r_v$. We then have
\begin{equation}    \label{c22}
    M_{\rm p}({R_{\rm e}}) = M_{\rm p}({R}_{\rm cut})/2 \ .
\end{equation}
Figure~\ref{aec} shows the effective radius,
calculated numerically from equations (\ref{c21}) and (\ref{c22}).
For $\widetilde R_{\rm cut} = 1$, a useful approximation, good to better
than 2\% relative accuracy, is:
{\samepage
\begin{eqnarray}
    R_{\rm e} / r_v &=& 0.5565 - 0.1941\,\log c \nonumber \\
    & & \mbox{} - 0.0756\,\log^2 c + 0.0331\,\log^3 c \ .
\end{eqnarray}
}

The prediction for the surface brightness $I = \Sigma_M/\Upsilon$ with
$\Sigma_M$ given by equation (\ref{c20}) expressed in terms of
the effective radius and the corresponding effective brightness $I_{\rm e}
= I(R_{\rm e})$ is shown in the upper panel of Figure~\ref{de} for
different values of the concentration parameter $c$. For comparison, we
also show the de Vaucouleurs (1948) $R^{1/4}$ law describing the observed
surface brightness distribution in giant elliptical galaxies:
\begin{equation}    \label{c23}
    I(R) = I_{\rm e} \exp\{-b\,[(R/R_{\rm e})^{1/4}-1]\} \ ,
\end{equation}
where $b = 7.67$. Clearly, the NFW surface brightness profiles are poorly
fitted by the $R^{1/4}$ law, when using $R_{\rm cut} = r_v$ to define the
effective radius of the NFW profile.

\begin{figure*}
\begin{center}
    \leavevmode
    \setbox100=\hbox{\epsfxsize=12cm\epsffile{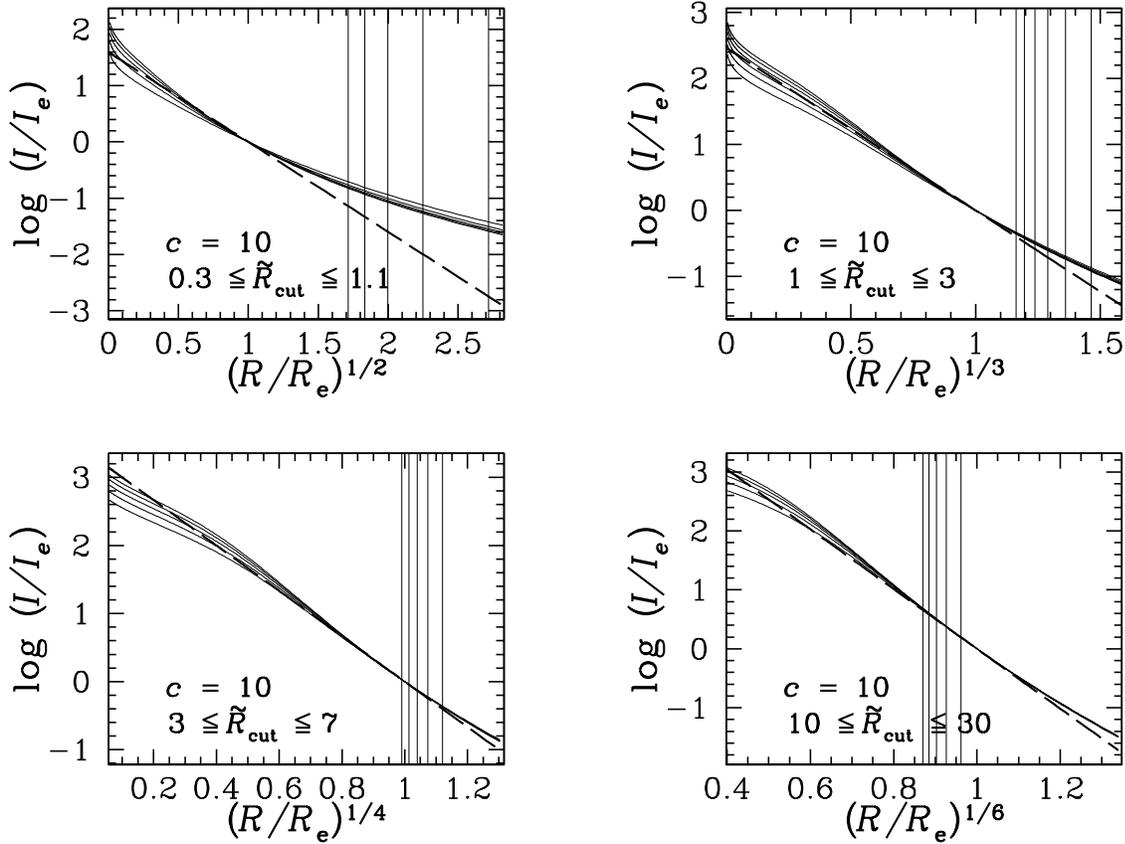}}
    \setbox101=\vbox{\rotr{100}}
    \hbox{\box101}
\end{center}
    \caption{Comparisons of $c=10$ projected NFW models
    (using eq.~[\ref{c20}]) to S\'ersic models (eq.~[\ref{c31}]).
    The curves represent the NFW models
    (for equally spaced values of $\widetilde R_{\rm cut}$ within
    the interval indicated in each plot,
    with $\widetilde R_{\rm cut}$ increasing upwards on the left portion
    of each plot). The S\'ersic law is shown as long dashed lines.
    The vertical lines represent the virial radius (with
    $\widetilde R_{\rm cut}$ increasing from right to left). }
\label{sersic10}
\end{figure*}

The lower panel of Figure~\ref{de} shows how the results depend on the
choice of cut-off for $c=10$ and $\widetilde R_{\rm cut} = 3$, 3.5, 4,
4.5, and 5. At first glance, it seems that the NFW profile is well
fitted by the $R^{1/4}$ law, especially for $\widetilde R_{\rm cut} \simeq
4$. However, the range of surface mass densities where the fit is
excellent is roughly $10^2$, and the fit is adequate for a range smaller
than $10^3$. In contrast, the surface brightness profile of the nearby
giant elliptical galaxy NGC~3379 (M~105) follows the $R^{1/4}$ law in a
range of 10 magnitudes (de Vaucouleurs \& Capaccioli 1979), i.e. a factor
$10^4$ in intensity.

In order to see how good is the de Vaucouleurs's fit in this case in both
panels of Figure~\ref{de} we plotted a number of data points equally
spaced in $R^{1/4}$. Since de Vaucouleurs \& Capaccioli (1979) do not
provide the error bars for their data, the error bars shown in the Figure
were taken from Goudfrooij et al. (1994). The excess of the data above the
$R^{1/4}$ law for small $R$ was already noted by de Vaucouleurs \&
Capaccioli (1979). The error bars are negligible for $R < R_{\rm e}$ and
smaller than 15\% out to $2.5 R_{\rm e}$, the maximum distance from the
centre reached in the data of Goudfrooij et al. (1994).

According to de Vaucouleurs \& Capaccioli (1979), in this galaxy the
$R^{1/4}$ surface brightness profile  extends to $R_{\rm lim} =
7.5\,R_{\rm e} = 26.4\,\rm kpc$, given a distance of 12.4 Mpc to NGC 3379
(Salaris \& Cassisi 1998). Within $R_{\rm lim}$, de Vaucouleurs \&
Capaccioli (1979) report a blue magnitude, corrected for galactic
extinction of $B = 10.10$, yielding a total blue luminosity of $2.2\times
10^{10}\,L_\odot$, hence a blue luminosity density of $2.8 \times 10^5
\,L_\odot\,\rm kpc^{-3}$. Since the mass within $R_{\rm lim}$ must be
greater than the mass in stars, we infer that within this radius,
$\Upsilon_B > 8$ (the typical mass to blue luminosity ratio for old
stellar populations), yielding an overdensity of the galaxy, relative to
the critical density $\rho_c$ of $v > 1.6\times 10^4/(H_0/70\,\rm km
\,s^{-1} \, Mpc^{-1})^2$. Therefore, since $v \gg 100$ (the value at
$r_v$), we conclude that $R_{\rm lim} \ll r_v$, hence $R_{\rm e} \ll
r_v/7.5$. In contrast,  with $\widetilde R_{\rm cut} = 1$, the effective
radius of the NFW model ($c=10$) is $\simeq 0.3\,r_v$ (Figure~\ref{aec}).
This discrepancy in $R_{\rm e}/r_v$ between NFW and $R^{1/4}$ law gets
even worse if one adopts $\widetilde R_{\rm cut} = 4$, which provides the
best fits of the NFW surface mass density to the $R^{1/4}$ law: indeed,
Figure~\ref{aec} indicates $R_{\rm e} \simeq 0.8 \,r_v$
for the NFW model.

In summary, the NFW surface mass density profile resembles an $R^{1/4}$
law in a fairly wide range of radii, but 1) one has to resort to an
abnormally large effective radius, very close to the virial radius, and
assume that the effective radius measures half the projected light (or
mass) within 4 times the virial radius, and 2) the fit is good in
a considerably
smaller range of radii than is observed in the nearby giant elliptical
NGC~3379.

The  generalization of the $R^{1/4}$ law into an $R^{1/m}$ law, first
proposed by S\'ersic (1968), is known to fit the surface brightness profiles
of elliptical galaxies within a much larger mass range than the de
Vaucouleurs law (Caon, Capaccioli \& D'Onofrio 1993). The surface brightness
of the S\'ersic profile is
\begin{equation}    \label{c31}
    I(R) = I_{\rm e} \exp\{-b(m)\,[(R/R_{\rm e})^{1/m}-1]\} \ ,
\end{equation}
where $b(m)$ is tabulated by Ciotti (1991), who gives the empirical relation
$b(m) \simeq 2\,m-0.324$, good to 0.1\% relative accuracy. The de
Vaucouleurs law is reproduced for $m=4$, while $m=1$ corresponds
to an exponential law as in spiral disks.

In Figure~\ref{sersic10}, we plot the NFW surface
brightness $I =\Sigma_M/\Upsilon$, with $\Sigma_M$ given by
equation~(\ref{c20}) and $c=10$, as a function of $(R/R_{\rm e})^{1/m}$
for various values of the S\'ersic parameter $m$. We compare them to the
S\'ersic profiles given by the straight dashed lines. The agreement is
good for all values of $m$, within ranges of $I/I_{\rm e}$ that increase
with increasing $m$. Comparison of the plots for different $m$ shows that
the S\'ersic models with lower $m$ generally agree better with the NFW
surface brightness for smaller radii, while those with larger $m$ are in
better agreement at larger radii, closer to the virial radius. Overall,
the NFW profile matches best the $m=3$ S\'ersic law, over a factor of
$10^3$ in intensity (7.5 magnitudes).

\begin{figure}
\begin{center}
    \leavevmode
    \epsfxsize=8cm
    \epsfbox[50 50 340 310]{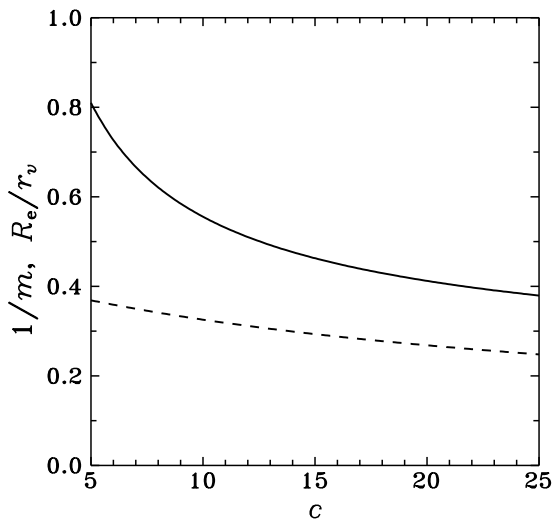}
\end{center}
    \caption{The best fitting parameters of the S\'ersic law, eq.
    (\ref{c31}), as functions of concentration: $1/m$ (dashed line) and
    $R_{\rm e}/r_v$ (solid line).}
\label{fitser}
\end{figure}

\begin{figure}
\begin{center}
    \leavevmode
    \epsfxsize=8cm
    \epsfbox[50 50 340 310]{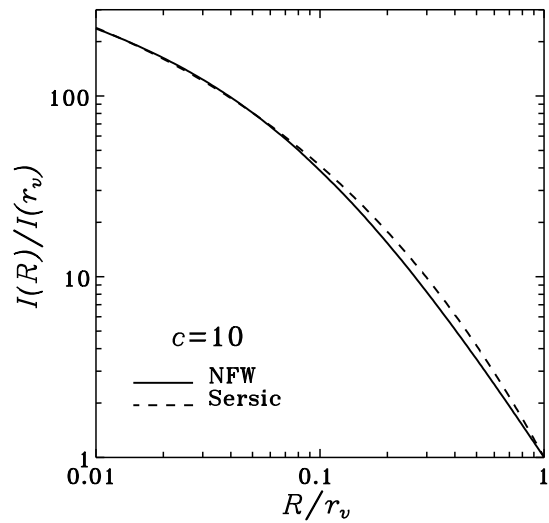}
\end{center}
    \caption{Comparison of the projected NFW density profile, eq.
    (\ref{c20}), and the best fitting S\'ersic law, eq. (\ref{c31}).}
\label{nfwser}
\end{figure}

For a more quantitative comparison, we performed
two-parameter fits of the S\'ersic models (\ref{c31}) to the projected NFW
formula (\ref{c20}). The NFW profile was sampled in the range of $0.01 <
\widetilde R < 1$ with a given $c$. The fitted S\'ersic parameters
$1/m$ and $R_{\rm e}/r_v$ obtained for different $c$ are shown in
Figure~\ref{fitser}. Figure~\ref{nfwser} compares the two projected
profiles for $c=10$. The best-fit parameters of the S\'ersic model in
this case are $m=3.07$ and $R_{\rm e}/r_v=0.55$.

While Caon et al. (1993) find similar ranges of agreement between observed
profiles and S\'ersic laws, this range in intensity is still smaller than
the range of $10^4$ found for NGC~3379 by de Vaucouleurs \& Capaccioli
(1979). Moreover, while Caon et al. (1993) find that the best fitting
S\'ersic models for elliptical galaxies have indices spanning a wide
range, from $m=2$ for faint ellipticals to $m=10$ for bright ellipticals,
the S\'ersic laws that match the NFW models span a much smaller range,
roughly $m = 3\pm0.5$ ($2.71 < m < 3.41$ for $5 < c < 15$, see
Figure~\ref{fitser}). Moreover, the problem of very high values of
$R_{\rm e}/r_v$ ($0.46 < R_{\rm e}/r_v < 0.81$ for $5 < c < 15$, see
Figure~\ref{fitser}), remains in the fits of S\'ersic profiles to
projected NFW models.

\section{Discussion}

The main disadvantage of the NFW model is the logarithmic divergence of
its mass (and luminosity for constant mass-to-light ratio). In
contrast, the Jaffe (1983) and Hernquist (1990) models converge in mass,
and their properties can be expressed in units of their asymptotic mass.
For the NFW model, one is restricted to a mass at a physical radius such
as the virial radius. This mass divergence also complicates the analysis
of surface brightness profiles, which involve the effective radius where
the aperture luminosity is half its asymptotic value. However,
independently of the radial cut-off introduced to define the effective
radius, the projected NFW density profile is consistent with constant
mass-to-light ratio, given the observed S\'ersic profiles of elliptical
galaxies, but only in a limited range of radii, with unusually high values
of $R_{\rm e}$ and in a smaller interval of S\'ersic shape parameters than
observed. On the other hand, the Hernquist (1990) model, whose density
profile scales as $r^{-4}$ at large radii, produces better fits to the
$R^{1/4}$ law.

The upper panel of Figure~\ref{de} suggests that, for reasonable
effective radii, if indeed dark matter follows the NFW profile, the
mass-to-light ratio, $\Upsilon$, is not constant but increases with radius,
not only in the outer regions, as is inferred from the
commonly accepted picture of galaxies embedded in more spatially extended
dark haloes, but also in the inner regions. This is at odds with the
observed kinematics of ellipticals that Bertola et al. (1993) inferred
from observations of ionised and neutral gas around specific ellipticals.
Moreover, increasing $\Upsilon$ throughout the galaxy implies radial
velocity anisotropy throughout elliptical galaxies, whereas violent
relaxation should cause isotropic cores.\footnote{Note that recent, state
of the art observations and modelling by Saglia et al. (2000) and Gebhardt
et al. (2000) do not strongly constrain the gravitational potentials of
elliptical galaxies, although NFW potentials may turn out to be
inconsistent with the current data. On the other hand, Kronawitter et al.
(2000) are able to rule out constant $\Upsilon$ for some elliptical
galaxies.} Thus it appears difficult to reconcile the photometry and
kinematics of elliptical galaxies with NFW models. In a forthcoming paper
(Mamon \& {\L}okas, in preparation), we will omit the assumption of mass
follows light in a more detailed assessment of the compatibility of the
observations of elliptical galaxies with the NFW model.

The results presented in this paper can be directly applied to the
analysis of the mass and light distribution in clusters of galaxies. A
standard procedure to do it is to measure the surface brightness and the
light-of-sight velocity dispersion and assuming some form of velocity
distribution or mass-to-light ratio calculate the luminosity density and
the velocity dispersion by solving the Abel integral equations (\ref{c20})
and (\ref{c24}) and the Jeans equation (Binney \& Mamon 1982, Tonry 1983,
Solanes \& Salvador-Sol\'{e} 1990, Dejonghe \& Merritt 1992). The
results of this procedure are uncertain because it involves derivatives of
observed quantities which are usually noisy. One also experiences a
degeneracy because different models fit the data equally well
(Merritt 1987). Instead of solving the Abel equations one can also model
the luminosity density and velocity dispersion with simple functions and
fit their parameters so that they reproduce their projected counterparts
(Carlberg et al. 1997).

Our results are useful for the simpler approach of assuming realistic
forms of the density distribution, velocity distribution and mass-to-light
ratio. Here we provide the tools for modelling the NFW density profile
with different velocity distributions and constant
mass-to-light ratio ($\Upsilon={\rm const}$), and obtain exact predictions
for the surface brightness and the line-of-sight as well as aperture
velocity dispersion that can be directly compared to observations.

\section*{Acknowledgements}

We thank Daniel Gerbal and Bernard Fort for useful conversations, and an
anonymous referee for helpful comments. EL\L \
acknowledges hospitality of Institut d'Astrophysique de Paris, where
part of this work was done. This research was partially supported by the
Polish State Committee for Scientific Research grant No. 2P03D00813 and
2P03D02319 as well as the Jumelage program Astronomie France Pologne of
CNRS/PAN.

\end{document}